\documentclass[twocolumn,superscriptaddress]{revtex4}
\usepackage{graphicx}

\topmargin by 2 \advance \topmargin -1.5in \leftmargin 8.5in
\advance \leftmargin -\textwidth \divide \leftmargin by 2 \advance
\leftmargin

\newcommand{\ave}[1]{\langle #1 \rangle}

\begin{document}
\author{Nima Sarshar}
\email{nima@ee.ucla.edu}
\affiliation{Department of Electrical Engineering, University of California,
Los Angeles}
\author{P. Oscar Boykin}
\email{boykin@ece.ufl.edu} \homepage[WWW:
]{http://boykin.acis.ufl.edu/} \affiliation{Department of
Electrical and Computer Engineering, University of Florida}
\author{Vwani Roychowdhury}
\email{vwani@ee.ucla.edu}
\affiliation{Department of Electrical Engineering, University of California,
Los Angeles}
\title{Finite Percolation at a Multiple of the Threshold}

\begin{abstract}
Bond percolation on infinite heavy-tailed power-law random
networks lacks a proper phase transition; or one may say, there is
a phase transition at {\em zero percolation probability}.
Nevertheless, a finite size percolation threshold $q_c(N)$, where
$N$ is the network size, can be defined. For such heavy-tailed
networks, one can choose a percolation probability $q(N)=\rho
q_c(N)$ such that $\displaystyle \lim_{N\rightarrow
\infty}(q-q_c(N)) =0$, and yet $\rho$ is arbitrarily large (such a
scenario does not exist for networks with non-zero percolation
threshold). We find that the critical behavior of random power-law
networks is best described in terms of $\rho$ as the order
parameter, rather than $q$. This paper makes the notion of the
phase transition of the size of the largest connected component at
$\rho=1$ precise. In particular, using a generating function based
approach, we show that for $\rho>1$, and the power-law exponent,
$2\leq \tau<3$, the largest connected component scales as $\sim
N^{1-1/\tau}$, while for $0<\rho<1$ the scaling is  $\sim
N^{\frac{2-\tau}{\tau}}$; here, the maximum degree of any node,
$k_{max}$, has been assumed to scale as $~N^{1/\tau}$. In general,
our  approach yields that for large $N$, $\rho \gg 1$, $2\leq
\tau<3$, and $k_{max} \sim N^{1/\tau}$, the largest connected
component scales as $\sim \rho^{1/(3-\tau)}N^{1-1/\tau}$.Thus, for
any fixed but large N, we recover, and make it precise, a recent
result that computed a scaling behavior of $q^{1/(3-\tau)}$ for
``small $q$''. We also provide large-scale simulation results
validating some of these scaling predictions, and discuss
applications of these scaling results to supporting efficient
unstructured queries in peer-to-peer networks.

\end{abstract}
\maketitle

\section{Introduction}
Percolation on random graphs with heavy tailed power-law degree
distributions is known to possess certain unique properties. Unlike
most cases of finite and infinite dimensional percolation, the
percolation on an infinitely large, heavy tailed power-law (PL)
random graph lacks a percolation threshold, i.e., the phase transition
happens at zero percolation probability. This is the source of
many interesting critical behavior in heavy-tailed random PL
networks that are remarkably different from those observed in networks with
non-zero percolation thresholds. In this paper,  we are particularly interested in the
scaling of the size of the largest connected component close to the
criticality, for networks of
finite but large size $N$. To that end, we first make the notion of closeness
to criticality precise, and provide a summary of our results.

\subsection{Summary of the Results}

 Consider a heavy-tailed PL network with
exponent $2\leq \tau<3$. A finite size percolation threshold
$q_c(N)$ can be defined for these networks that goes to zero as
$N$ goes to infinity.

 Consider percolation at a probability
$q(N)=\rho q_c(N)$. After bond percolation with probability
$q(N)$, a total of around ${\cal H}(N,\rho)\triangleq q(N) N
\langle k \rangle/2$ links remain in the network, where $\langle k
\rangle$ is the average degree. Of these ${\cal H}(N,\rho)$ links,
a number ${\cal G}(N,\rho)$ of them belong to the same largest
connected component. This paper answers the following questions:
How does ${\cal G}(N,\rho)$ scale with $N$ and $\rho$?  How does
the fraction ${\cal G}(N,\rho)/{\cal H}(N,\rho)$ scale with $N$
and $\rho$? In other words, what is the chance that a link left
after the percolation belongs to the giant connected component?
Does this probability converge to some constant?

 The answers to these questions are known for
networks with finite percolation threshold $q_c=q_c(\infty)>0$ (
note that $\rho$ is meaningful only when $\rho\leq 1/q_c$,
otherwise one has $q>1$). In particular, ${\cal G}(N,\rho)\sim N$
for $1<\rho\leq 1/q_c$, and ${\cal G}(N,\rho)=O(k_{max} \log N)$
for $0<\rho<1$, where $k_{max}$ is the largest degree of the nodes
in the network. When $\rho=1$, ${\cal G}(N,\rho)$ is known to
scale as $N^{2/3}$ \cite{N01} for most random networks under
certain constraints. The probability that a random link after
percolation belongs to the largest connected component, ${\cal
G}(N,\rho)/{\cal H}(N,\rho)$, converges to a constant
$\omega(\rho)>0$ when $1/q_c\geq \rho>1$, and goes to zero at
least as fast as  $\frac{k_{max}\log N}{N}$ when $0\leq \rho<1$.

\begin{figure}
\begin{center}
\includegraphics[width=2.5in,height=2.75in]{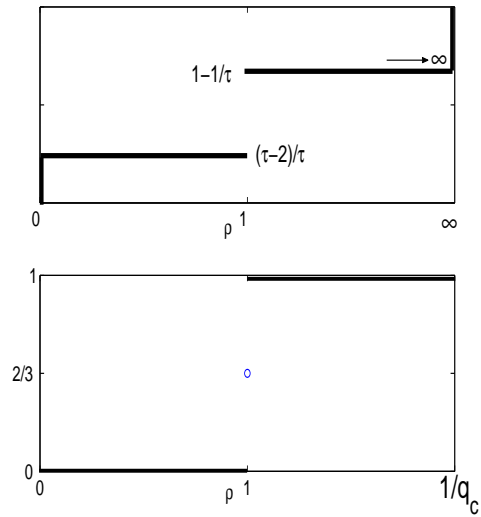}
\caption{The scaling exponent, $\lambda_\tau(\rho),$ undergoes a first order phase transition {\em for all} random networks at $\rho=1$. Recall that $\lambda_\tau(\rho)=\lim_{N\rightarrow
\infty} \log({\cal G}(N,\rho))/\log(N)$, where ${\cal G}(N,\rho)$ is the size of the
largest connected component in a network of size $N$ after bond percolation with probability $\rho q_c$.  The top figure
shows the transition for a heavy-tailed power-law random network
while the bottom figure corresponds to a random network with a
bounded $k_{max},$ and therefore with non-zero percolation
threshold.}\label{fig:res}
\end{center}
\end{figure}

The answers to the above questions for heavy-tailed PL networks
turn out to be distinctly different from those for random networks
with finite percolation thresholds. In order to capture these
differences succinctly, we define a {\em scaling exponent},
$\lambda_\tau(\rho)\triangleq\lim_{N\rightarrow \infty} \log({\cal
G}(N,\rho))/\log(N)$. Fig. \ref{fig:res} shows a schematic of some
of the results that we derive, based on our application of the
generating function approach.

 For $1< \rho \ll N^{1/\tau}$, we
show that ${\cal G}(N,\rho)\sim \langle k \rangle N^{1-1/\tau}$
(see Eqns. \ref{eqn:g0},~\ref{eqn:g1}). In other words, the
scaling exponent is $\lambda_\tau(\rho)=1-1/\tau$ for any
$\rho>1$, and thus, the size of the giant connected component does
not scale linearly with $N$ for constant $\rho>1$.

For $0<\rho <1$, the scaling exponent $\lambda_\tau(\rho) =
1-2/\tau$. This last statement follows from the observation that
after percolation, the expected degree of a node with degree
$k_{max}$ will  reduce to $\rho q_ck_{max}$. This last statement
follows from the observation that after percolation, a node with
degree $k_{max}$ will have an average degree of $\rho q_c
k_{max}$. Taking $k_{max}\sim N^{1/\tau}$ and using the fact that
$q_c \sim N^{1-3/\tau}$ (see Eqn. \ref{eqn:qc1}) gives the desired
result.

 Therefore, {\em
if we consider $\rho$ as the order parameter}, then there is a
{\em phase transition} at  $\rho=1$ for  {\em all random networks}
(irrespective of whether it is heavy-tailed with unbounded
variance or not) except that the phase transition levels are
different (Fig. \ref{fig:res}). This provides a unified view of
the phase transition phenomenon for largest connected components
in random networks.

The above result enables us to calculate the following interesting
quantity; ${\cal G}(N,\rho)/{\cal H}(N,\rho)$ (see Eqn.
\ref{eq-dpcase1}). We find that for $\rho>1$, the ratio ${\cal
G}(N,\rho)/{\cal H}(N,\rho)\sim N^{(\tau-2)/\tau}$ for $2\leq
\tau<3$. In other words, {\em the probability that a randomly
chosen link} after the percolation {\em belongs to the giant
connected component  goes to zero} as $N^{(\tau-2)/\tau}$ for
$2<\tau<3$. Only for $\tau=2$, a finite fraction of all the links
that remain after the percolation belong to the giant connected
component.

\subsection{Relation to Previous Work}

The finite-size scaling of the cardinality of the largest
percolation components close to criticality has always been a
source of interest. While for finite dimensional percolation
problems, the issue is still subject to a great deal of
controversy \cite{Gri99}, it has been successfully resolved for
percolation on most random graphs (that are examples of infinite
dimensional percolation). In particular, for an Erdos-Reyni random
graph with percolation threshold $q_c$, the size of the largest
connected component is known to scale as $\Theta(N^{2/3})$ when
$q=q_c+O(N^{-2/3})$. Thus, right at the percolation threshold
$q_c$, the size of the largest connected component scales with the
network size as $\sim N^{2/3}$ \cite{Bollobas01}. Furthermore, it
can be shown that for any fixed large $N$, the size of the largest
connected component scales linearly as $|q-q_c|$ when $q-q_c$ is
small. This indicates, among other things, the continuity of the
size of the largest connected component as a function of $q$ at
$q=q_c$. Note that in such cases, the percolation threshold $q_c$
is finite, and independent of $N$. In \cite{N01}, this result has
been extended to random networks on any given degree distribution,
subject to certain constraints.

Recent results in \cite{new} have shown that critical scaling
properties of random power-law networks with exponents $\tau$ in
the range $[2,4]$, are significantly different from those of most
other random networks. For heavy-tailed PL networks (i.e., $2\leq
\tau<3$), which is the case of  interest in this paper, results in
\cite{new} suggest that the size of the largest connected
component scales as $|q-q_c|^{1/(3-\tau)}$ for any fixed but large
$N$ (as opposed to a linear scaling of $|q-q_c|$ for most other random
networks). The approach used in this work is also based on the
generating function approach adopted in this paper. Our goal, however,
is to study the scaling laws as a function of $N$ for a fixed $q/q_c$.

We first argue that the relevant order parameter in
heavy-tailed power-law random networks is $\rho=q/q_c$ and
critical percolation properties are best explained in terms of
$\rho$. We then derive, in a unified way, the scaling of the size
of the largest connected component as a function of $\rho$ {\em
and} $N$. We show that for any fixed $\rho>1$, the size of the
largest connected component scales as $N^{1-1/\tau}$, in contrast
to the universal scaling of $O(N^{2/3})$ for most random networks
with finite size percolation threshold. Interestingly, for $\tau$
approaching $3$, where the network assumes a finite size
percolation threshold, the universal scaling exponent $2/3$ is
recovered. On the other hand, for $\rho<1$, the size of the
largest connected component scales at most as fast as
$N^{(\tau-2)/\tau}$. For fixed but large $N$, we find that the
size of the largest component as a function of $\rho$ for all
$\rho$, and we recover the results in \cite{new} for $1 \ll \rho
\ll N^{1/\tau}$. In general, for large $\rho \gg 1$ and $N$, the
size of the largest connected component is shown to scale as $\sim
\rho^{1/(3-\tau)}N^{1-1/\tau}$.

\subsection{Applications}
The results in this paper also have applications to  peer-to-peer
(P2P) data mining or search algorithms on unstructured complex
networks. Percolation at a multiple of the threshold is at the
heart of a scalable P2P search algorithm, called percolation
search algorithm, introduced by the authors in \cite{perc}.
Percolation search uses a probabilistic broadcast algorithm to
reduce the number of communications necessary for finding contents
in a large scale complex network that has a heavy-tailed power-law
degree distribution. The percolation probability $q$ will
correspond to the probability with which nodes of the network
communicate a message to any of their neighbors. For the search
algorithm to work efficiently, one needs to choose a value of $q$
that is as small as possible and yet results in a  large enough
connected component; that is, choose $q$ to be a constant multiple
of the threshold $q_c$. The scaling of this connected component
directly specifies the scaling of the P2P search traffic with
network size.  The results in this paper form the theoretical
basis for showing that this traffic scales sub-linearly with the
network size for any heavy-tailed random power-law network.

The rest of this paper is organized as follows: In Section
\ref{sec:gen} we briefly review the generating function formalism
\cite{N01} for calculating the percolation properties of random
networks on given degree distributions. We then specialize the
approach to the case of percolation at a multiple of the threshold
for heavy-tailed random power-law networks and derive the scaling
relation of the size of the connected component as a function of
both $q/q_c$ and the network size. A number of simulations are
reported in Section \ref{sec:sim} which verify our results.
Concluding remarks are made in Section \ref{sec:conc}.

\section{Calculation of the Cluster Sizes}\label{sec:gen}

In this section, we introduce the analytic approach of the paper.

\subsection{Modeling Finite Size Effects}

Consider a random graph with degree distribution $p_k$,
$k=1,2,...,k_{max}$. Thus, the probability that a randomly chosen
node of this graph has degree $k$ is $p_k$, where $1\leq k\leq k_{max}$.
 We
assume that the maximum degree is such that \emph{every} network
is expected to have at least one node of that degree: $N
p_{k_{max}} \ge 1$. For a power-law random graph $p_k = A_{\tau}
k^{-\tau}$ for $k\in (1, k_{max})$, we have:
\begin{eqnarray}
A_{\tau} &\approx& \zeta(\tau)\\
k_{max} &\approx& \frac{N^{1/\tau}}{A_{\tau}^{1/\tau}}
\label{eqn:kmax}
\end{eqnarray}
This choice of $k_{max}$ reflects the maximum scaling of $k_{max}$
that results in the expected number of high degree nodes to be
only one.

Thus,  as $N$ increases , $k_{max}$ will increase and the variance
of the distribution will become unbounded for $\tau<3$. We focus
on the parameter regime, $2\leq \tau <3$, where the networks have
degree distributions with unbounded variance but
 bounded mean (except at $\tau=2$, where it increases only logarithmically).
Though we are interested in finite size effects, we focus on the case of large
values of $N$ where $\ave{k^2} \gg \ave{k}$.

An accompanying assumption we make is that the generating
function approach, which is meant to deal with infinite size
networks, is still applicable to random power-law networks of
finite, but large size, $N$. The same assumption is implicitly made in
\cite{new}, which also uses the same generating function formalism. A formal treatment of the validity of this assumption is beyond
the scope of this paper.  It is not, however, difficult to see that
the scaling of  the variance (and, hence, of $k_{max}$) with $N$ will play an important role
in any such formal approach, and for a discussion on choosing $k_{max} \sim N^{1/\tau}$, see \cite{MR95,Aiello00}. We have verified the predictions made by the generating function approach
via large-scale simulation results, some of which are presented in Section \ref{sec:sim} and in Appendix \ref{ap:perc}. The simulation results in Appendix \ref{ap:perc}
show that the predicted infinite-size percolation thresholds are matched
closely by numerical estimates obtained from synthetic finite-size
networks; the match is very good even for relatively small size networks.

\subsection{Percolation as a Branching Process}

 Just
above the criticality, the largest component of the network is a
tree. Percolation on a tree can be viewed as a branching process.
Using this fact, we will treat percolation as branching process
using the generating functions method introduced in \cite{N01}.
The generating function for the degree of a randomly chosen node
can be defined as $G_0(x)=\sum_{k=1}^{k_{max}}p_k x^{k}$. The
generating function for the degree of a node arrived at by
following one end of a randomly chosen link is
$G_1(x)=G'_0(x)/G'_0(1)$.

The bond percolation with probability $q$ on a given graph is as
follows: For each link of the network, delete the link with
probability $1-q$ and retain it with probability $q$,
independently.  The distribution of the size of the connected
components of a random graph after bond percolation with
probability $q$ can in principle be found. Throughout the rest of
the paper, we use results in \cite{CNSW00},\cite{N01}.

Let $u$ be the probability that a random link does not lead
to an infinite set of nodes, one then has:
\begin{equation}\label{eqn:const}
u = 1-q + qG_1(u)
\end{equation}

This equation is understood as follows: Take any random link. The
probability that the link is deleted (and thus does not lead to a
giant component) is $1-q$ (hence the first term on the right hand
side). Now follow the link to one of its random ends to arrive at
a node $V$. $G_1(x)$ is the generating function of the
distribution of the number of links that go out of $V$. Being a
random graph, the probability of each of those links to lead into
a giant connected component is again $u$.  The probability that
none of the remaining links of $V$ go to an infinite component is
$G_1(u)$. Hence we have (Eqn. \ref{eqn:const}).

As far as the nodes are concerned, the probability that a random
node does not lead to an infinite number of nodes is thus
$G_0(u)$, and hence the fraction of nodes in the infinitely large
component is:
\begin{equation}\label{eqS}
S=1 - G_0(u)
\end{equation}

Below the percolation threshold, the only solution to (Eqn.
\ref{eqn:const}) is $u=1$. In fact, this can be used to show that
the percolation threshold of any random graph on a given degree
distribution is given by:

\begin{equation}\label{eqn:qc}
q_c=G'_1(1)^{-1}= \frac{ \ave{k} }{ \ave{ k^2 } - \ave{k}}
\end{equation}

When working just above the percolation threshold (as is the
subject of this paper), one expects  that $u \lesssim 1$, and
hence we look for  solutions of the form $u=1-\delta$ with
$\delta\approx 0$:
\begin{eqnarray}\label{maineq}
u &=& 1-q + qG_1(u)\nonumber\\
1 - \delta &=& 1 - q + q G_1(1 - \delta)\nonumber\\
\frac{\delta}{q}&=& 1 - G_1(1-\delta)\nonumber\\
\end{eqnarray}

Follow a particular edge, define $\delta$ the probability that it
leads to an infinite component.  In order to solve for $\delta$,
one can write a Taylor series expansion of $G_1(1-\delta)$:
\begin{equation}
\label{eq:g1_series}
G_1(1-\delta)=\sum_{n=0}^\infty G^{(n)}_1(1)\frac{(-\delta)^n}{n!}
\end{equation}

For heavy tailed random graphs, all these moments (except for the
mean $G'_0(1)$) are large (in fact infinite as $N$ goes to
infinity). Nevertheless, for any finite $N$, these moments are
still finite and the expansions can be carried out.
In this work, we deal with heavy-tail power-law random graphs.
For these graphs we have
$\ave{k^{n+1}} \approx \left(\frac{n-\tau}{n+1-\tau}\right) k_{max}\ave{k^n}$.
We use this property to make the following approximation:
\begin{eqnarray*}
G^{(n)}_1(1) &=& \frac{ \ave{ \prod_{i=0}^n (k-n) } }{ \ave{k} }\\
             & \approx & \frac{ \ave{ k^{n+1} }}{ \ave{k} }
\end{eqnarray*}

Finally, by approximating the sum
as an integral we can get that for $n \ge 1$
\begin{equation}\label{eqn:mom}
 \ave{ k^{n+1} } \approx \frac{ A_\tau
k_{max}^{n+2-\tau} }{n+2-\tau}
\end{equation}

While these approximations may seem crude, we show in Section \ref{sec:sim}
that the predictions we make match the simulation results.

\subsection{Solving for the Scaling Exponent}

We can put (Eqn. \ref{eqn:mom}) into (Eqn. \ref{eq:g1_series}):
\begin{eqnarray*}
G_1(1-\delta)&=&\sum_{n=0}^\infty G^{(n)}_1(1)\frac{(-\delta)^n}{n!}\\
&\approx& \sum_{n=0}^\infty \frac{\langle k^{n+1} \rangle  }{\langle k\rangle}
\frac{(-\delta)^n}{n!}\\
&\approx& 1 + \sum_{n=1}^\infty \frac{A_\tau}{\langle
k\rangle}\frac{k_{max}^{n+2-\tau}}{ (n+2-\tau) }\frac{(-\delta)^n}{n!}\\
&=&1 + \frac{A_\tau k_{max}^{2-\tau}}{\langle k\rangle}
   \sum_{n=1}^\infty \frac{(-\delta k_{max})^n}{n!(n+2-\tau)}
\end{eqnarray*}

Defining a finite size percolation threshold like any other random
network as $q_c=1/G'_1(1)$, one gets:

\begin{eqnarray*}
\frac{A_\tau k_{max}^{2-\tau}}{\langle k\rangle}&\approx&
\frac{(3-\tau)\langle k^2\rangle}{k_{max}\langle k\rangle}\\
&\approx& \frac{3-\tau}{k_{max}q_c}
\end{eqnarray*}

which gives:

\begin{eqnarray*}
G_1(1-\delta)&\approx& 1 + \frac{3-\tau}{k_{max}q_c}
\sum_{n=1}^\infty \frac{(-\delta k_{max})^{n}}{n!(n+2-\tau)}
\end{eqnarray*}

Inserting back into (Eqn. \ref{maineq}):

\begin{eqnarray*}
\frac{\delta}{q}&=& 1 - G_1(1-\delta)\\
&\approx& -\frac{3-\tau}{k_{max}q_c}
\sum_{n=1}^\infty \frac{(-\delta k_{max})^{n}}{n!(n+2-\tau)}\\
-\frac{q_c}{q(3-\tau)} (\delta k_{max}) &\approx&
\sum_{n=1}^\infty \frac{(-\delta k_{max})^{n}}{n!(n+2-\tau)}
\end{eqnarray*}

For each constant $\tau$, define $\alpha = \frac{q_c}{q(3-\tau)}$
and $z=\delta k_{max}$ to get an equation that is independent of
$k_{max}$ in the limit of large  $k_{max}$:
\begin{equation}\label{eq:z_alpha}
  -\alpha z = \sum_{n=1}^\infty \frac{(-z)^{n}}{n!(n+2-\tau)}
\end{equation}

(Eqn. \ref{eq:z_alpha}) is our first result of this paper. It
clearly states that as far as $k_{max}$ is large, $z$ would be
independent of $k_{max}$ and hence $N$. Thus, the fraction of
links in the giant connected component will scale as $\delta=z
k_{max}^{-1}$ for some constant $z$ depending only on $\tau$ and
$q/q_c$.

The number of links in the largest connected component after
percolation, will therefore scale as:
\begin{equation}\label{eqn:g}
{\cal G}(\tau,\rho)~\triangleq ~ \langle k \rangle \delta N~=~
\langle k \rangle k_{max}^{-1} z N
\end{equation}
where $z$ depends only on $\tau, \rho$ and $k_{max}$ depends only
on $N,\tau$. The above relation is valid as far as the ratio
$\rho=q/q_c$ for
\begin{equation}\label{eqn:qc1}
 q_c \approx \langle k
\rangle /\langle k^2 \rangle \approx  \langle k \rangle
k_{max}^{\tau-3}=  \langle k \rangle N^{1-3/\tau}
\end{equation}

When $2<\tau<3$, (Eqn. \ref{eq:z_alpha}) can be simplified in
terms of incomplete gamma functions. The incomplete gamma function
is defined for $\Re(a)>0$ as:
\begin{equation}
\nonumber \Gamma(a,z) = \int_z^\infty x^{a-1} e^{-x} dx
\end{equation}
At $z=0$ it takes on the familiar value: $\Gamma(a,0)=\Gamma(a)$.
For all $z$, the following holds true:
\begin{equation}
\label{eq:gamma_recur} \Gamma(a,z)=\frac{1}{a}(\Gamma(a+1,z) -
e^{-z}z^a)
\end{equation}

The series expansion for $\Gamma(a,z)$ is as below:
\begin{equation}
\label{eq:gamma_series} \Gamma(a,z)=\Gamma(a) -
z^a\sum_{n=0}^\infty \frac{(-z)^n}{n!(n+a)}
\end{equation}

Or equivalently using (Eqn. \ref{eq:gamma_series}):
\begin{equation}
\sum_{n=1}^\infty\frac{ (-z)^n }{ n!(n+a) }
 = \frac{\Gamma(a) - \Gamma(a,z)}{z^a} - \frac{1}{a}
\end{equation}

Applying this identity to (Eqn. \ref{eq:z_alpha}) with $a=2 -
\tau$:
\begin{equation}
\label{eq:z_beta} -\alpha z = \frac{\Gamma(2-\tau) -
\Gamma(2-\tau,z)}{z^{2-\tau}} - \frac{1}{2-\tau}
\end{equation}
 Applying the above to (Eqn. \ref{eq:z_alpha}), with
$a=2 - \tau$, and setting $\beta =
\frac{q_c}{q}\frac{\tau-2}{3-\tau}$
\begin{equation}\label{eqn:tr}
\label{eq:z_beta} \beta z = \frac{\Gamma(3-\tau) -
\Gamma(3-\tau,z)}{z^{2-\tau}} + e^{-z} - 1
\end{equation}
when $2<\tau<3$.

To see how the scaling results of \cite{new} follow from
(\ref{eqn:tr}), lets consider the case $z\gg 1$. In this case,
$\Gamma(a,z)\rightarrow 0$, hence:
\begin{eqnarray*}
-\alpha z &\approx& z^{\tau - 2}\Gamma(2-\tau) - \frac{1}{2-\tau}\\
\beta z &\approx& (2 - \tau)z^{\tau - 2}\Gamma(2-\tau)  - 1\\
&=& z^{\tau - 2}\Gamma(3-\tau) - 1
\end{eqnarray*}

If $z^{\tau - 2}\Gamma(3-\tau)\gg 1$, we can neglect $1$ to obtain:
\begin{eqnarray*}
\beta z &\approx& z^{\tau - 2}\Gamma(3-\tau)\\
z &\approx& \left( \frac{\Gamma(3-\tau)}{\beta}\right)^{1/{(3-\tau)}}\\
&=& \left( \frac{q (3-\tau)\Gamma(3-\tau)}{q_c(\tau - 2)} \right)^{1/(3-\tau)}\\
&=& \rho^{1/(3-\tau)} \left( \frac{\Gamma(4-\tau)}{(\tau - 2)}
\right)^{1/(3-\tau)}
\end{eqnarray*}

In other words, the fraction of the links in the largest connected
component $\delta$ scales as:

\[
\delta =  k_{max}^{-1} z \approx \left(
\frac{\Gamma(4-\tau)}{(\tau - 2)} \right)^{1/(3-\tau)}
\rho^{1/(3-\tau)} N^{-1/\tau}
\]

when $1 \ll \rho \ll k_{max}$ and $2<\tau<3$ and when the scaling
of $k_{max}$ is chosen according to Eqn. \ref{eqn:kmax} (i.e.,
$N^{1/\tau}$). The number of links in the largest connected
component will therefore scale as

\begin{equation}\label{eqn:g0}
{\cal G}(\rho,\tau)\sim \rho^{1/(3-\tau)}N^{1-1/\tau}
\end{equation}

 Similar result
follows for $\tau=2$ (see Appendix \ref{ap:tau2}). In particular,
the fraction of links in the largest connected component can be
shown to scale as $\sim N^{-1/2} \rho \ln \rho$ and therefore, the
total number of links in the largest connected component scales as
\begin{equation}\label{eqn:g1}
{\cal G}(\rho,2)\sim \rho \ln \rho \sqrt{N} \ln N
\end{equation}

Another interesting regime is where $\rho\rightarrow 1^+$ or $z\ll
1$. This will correspond to percolation very close to the
threshold $q_c$. If $z\ll 1$ we can look at the first few terms of
(Eqn. \ref{eq:z_alpha}):
\begin{eqnarray*}
-\alpha z &=& \sum_{n=1}^\infty \frac{(-z)^{n}}{n!(n+2-\tau)}\\
 &\approx& -\frac{z}{3-\tau} + \frac{z^2}{2(4-\tau)}\\
\delta&=& zk_{max}^{-1}= 2\left(1 - \alpha(3-\tau)\right)\frac{4-\tau}{3-\tau}k_{max}^{-1}\\
&=& 2\left(1 -
\frac{q_c}{q}\right)\frac{4-\tau}{3-\tau}N^{-1/\tau}\\&=&2\frac{4-\tau}{3-\tau}\left((\rho-1)/\rho\right)N^{-1/\tau}
\end{eqnarray*}
In other words, the scaling with the order parameter is again
linear close to the finite size percolation threshold.

\subsection{Links in the Largest Component After Percolation}

The parameter $\delta$ is the fraction of \emph{total} links
before percolation that are in the largest connected component
after percolation. With percolation threshold being small, most of
the links are deleted during the percolation process. We are
interested in the probability that a link \emph{not deleted}
during the percolation process is part of the largest connected
component. Equivalently, what fraction of the links that remain
after percolation are part of the largest connected component?

To answer this, note that the number of links that remain after
the percolation is closely approximated by $${\cal H}(N,\rho)=\rho
q_c \langle k \rangle$$

 Using (Eqn. \ref{eqn:qc1}) to calculate $q_c$ and the scaling results for $\delta$, the probability that a link left
after the percolation is part of the largest connected component
can be calculated as follows.

\begin{equation}
\frac{{\cal G}(N,\rho)}{{\cal H}(N,\rho)}\sim\left\{
\begin{array}{l}
\rho \ln \rho\;\;\;\;\;\;\;\;\;\;\;\;\;\;\;\;\;\;\;\;\;\;\;\tau=2 \\
\rho^{1/(3-\tau)} N^{2/\tau-2}\;\;\;\;\;\;2<\tau<3
\end{array}
\right. \label{eq-dpcase1}
\end{equation}

Interestingly, for $\tau>2$, the fraction of links that are part
of the giant connected component constitute to an infinitesimally
small fraction of all links after the percolation. Only for
$\tau=2$ a finite fraction of all the links that remain after the
percolation belong to a single connected component.

\section{Simulations}\label{sec:sim}

\begin{figure*}
\begin{center}
\includegraphics[width=2.9in,height=3.6in,angle=270]{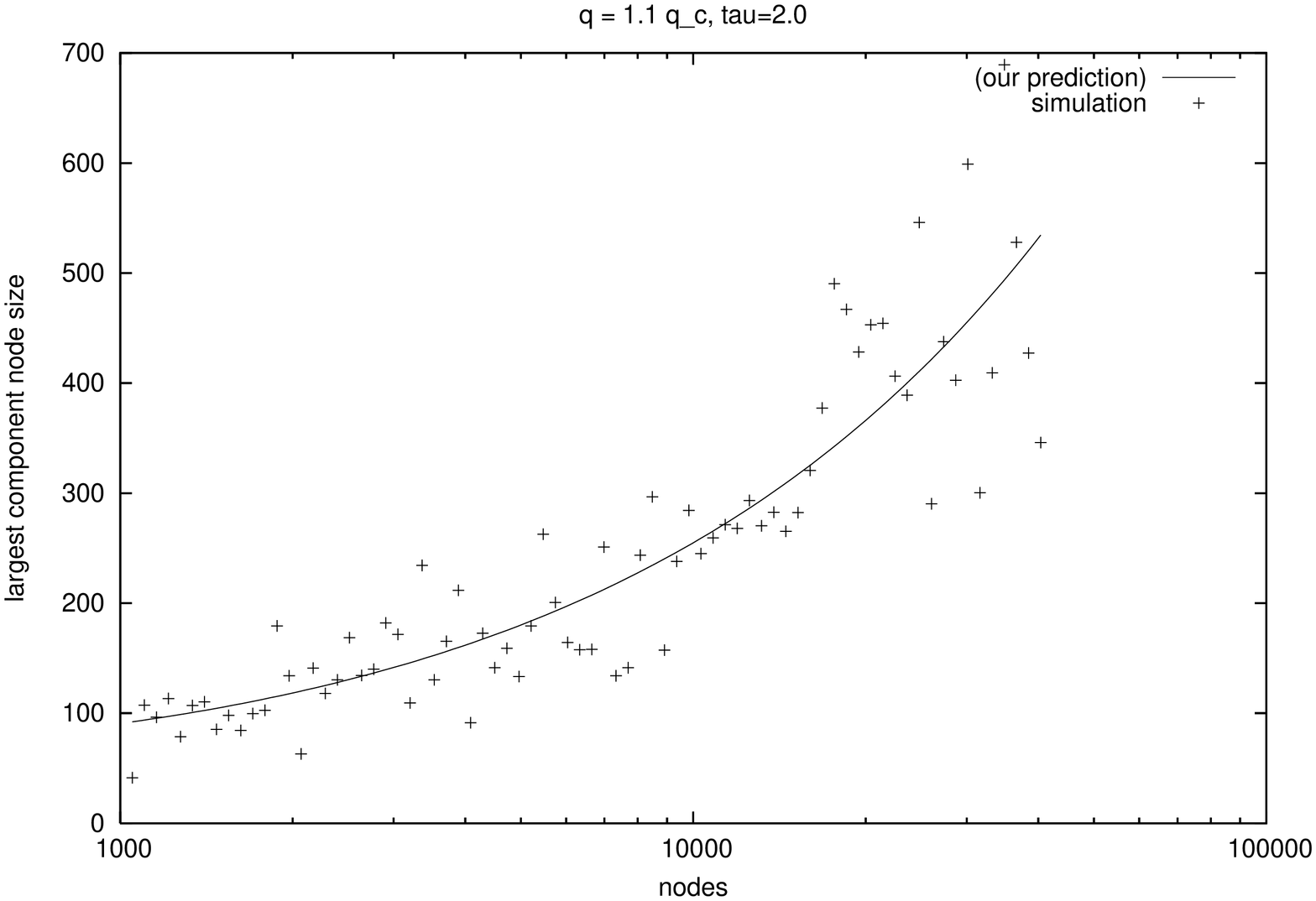}
\includegraphics[width=2.9in,height=3.6in,angle=270]{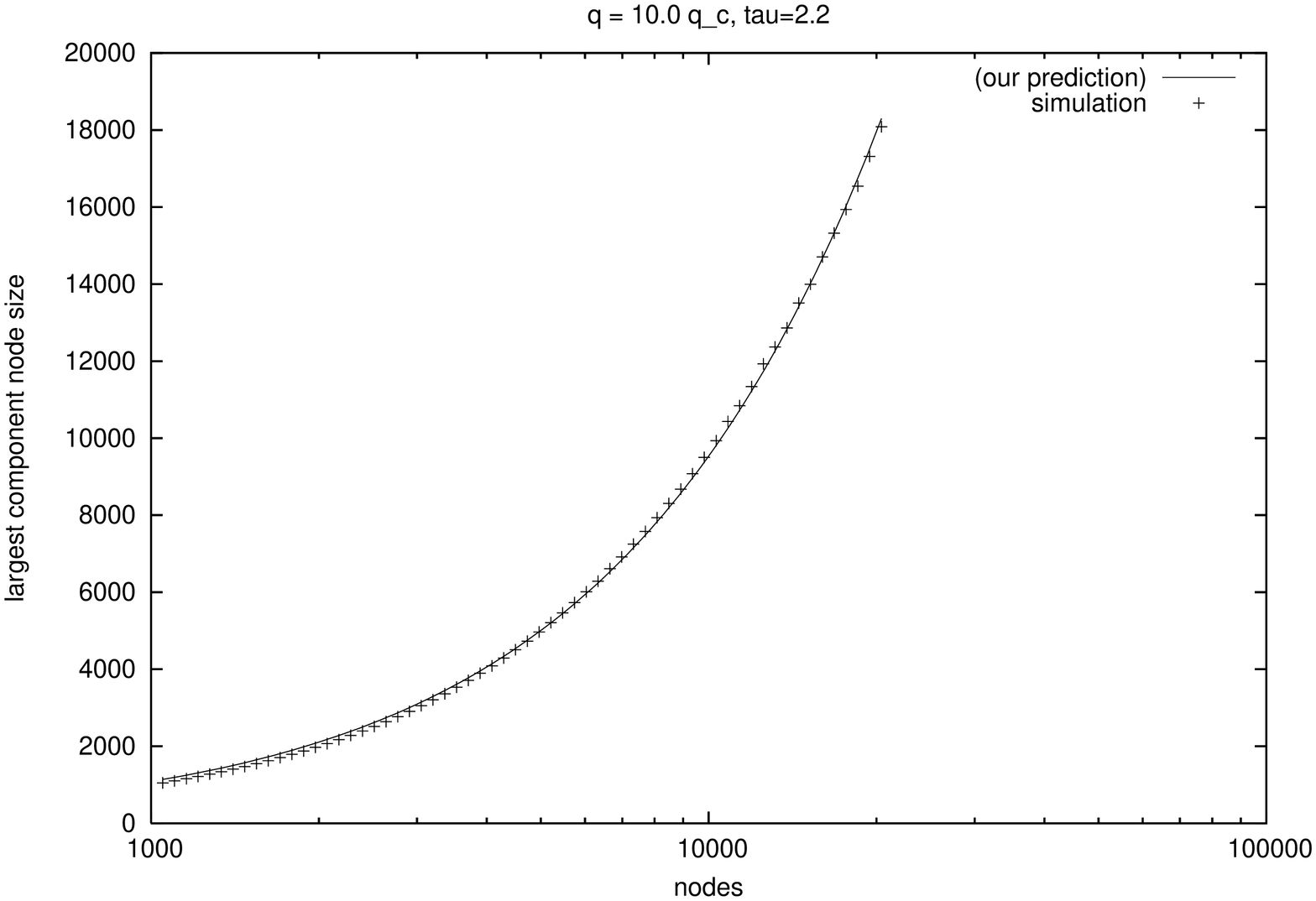}
\includegraphics[width=2.9in,height=3.6in,angle=270]{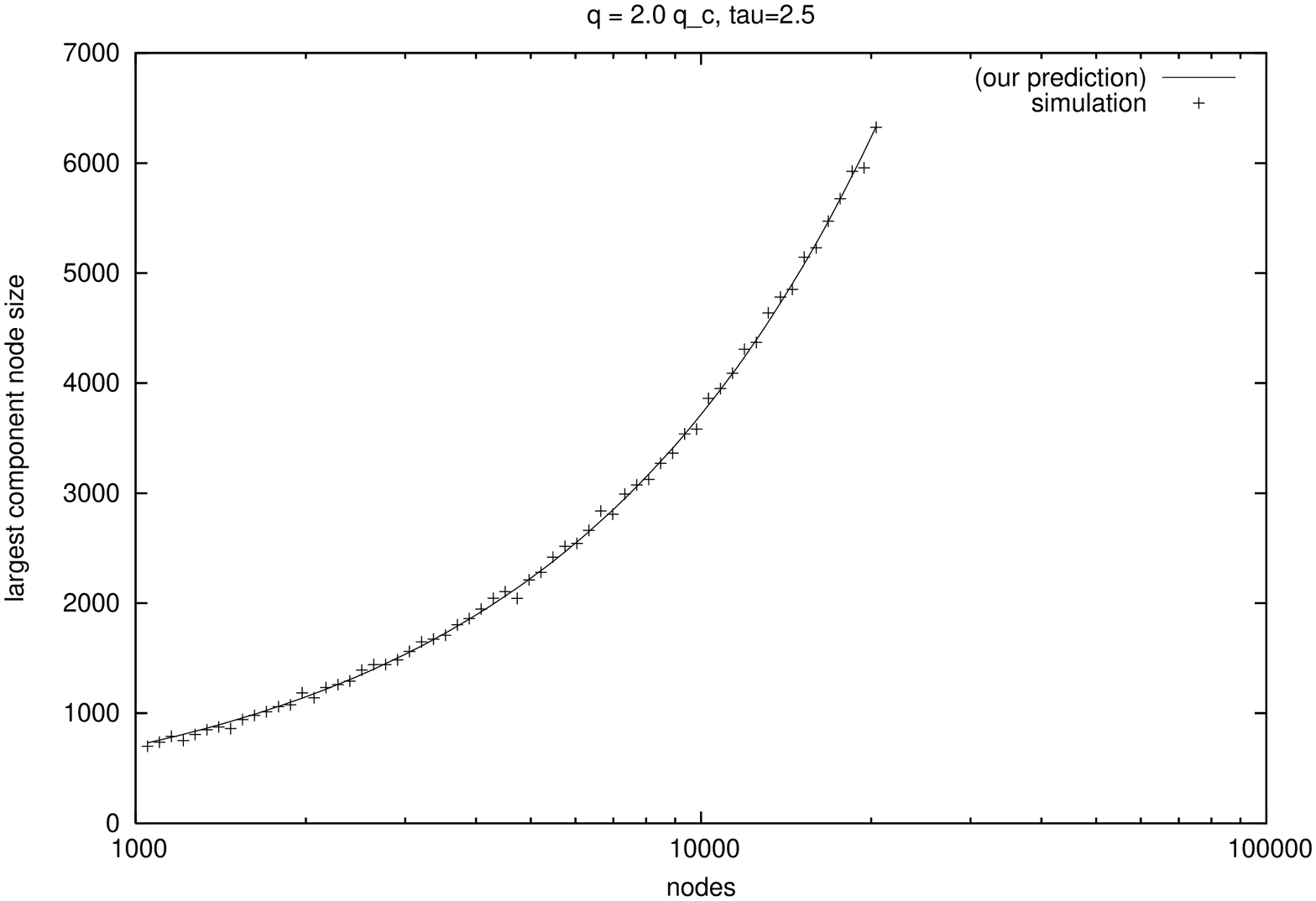}
\caption{Examples of the scaling of the size of the largest
connected component with the network size $N$ for different values
of $\tau$ and $q/q_c$: From top down, $\tau=2.0, q/q_c=1.1$,
$\tau=2.2, q/q_c=20.0$ and $\tau=2.5, q/q_c=10.0$. The scaling
$N^{1-1/\tau}$ is depicted on all the plots. Even in the case of
the top plot, where the largest component is as small as $100$ our
scaling predictions provide good matches to the simulations.}
\label{fig:scaling_N2}
\end{center}
\end{figure*}

Figure \ref{fig:scaling_N2} shows the scaling of the largest
component after bond percolation as a function of the size of the
original network $N$ when $q/q_c$ is a constant \footnote{The
source code of our simulator is available by contacting the
authors} using network sizes ranging from $1,000-1,000,000$ nodes
for different values of $\tau$. We compare the simulation result
to the scaling we would predict (i.e., $N^{1-\frac{1}{\tau}}$).
Simulation results suggest that our scaling law is correct as long
as $k_{max}$ is very large compared to $z$ and $q/q_c$ is in the
order of one.

\section{Concluding Remarks}\label{sec:conc}

We investigated the properties of the
size of the giant connected component just above the percolation
threshold in heavy-tailed power-law random graphs, for which the
percolation probability is known to be vanishingly small. By
normalizing the percolation probability by the percolation
threshold, we were able to trace the scaling behavior of the size of
the giant connected component at very small percolation
probabilities. In particular, we showed that $\delta$, the fraction of links in the giant
connected component close to the percolation threshold, is
proportional to the factor $\frac{1}{k_{max}}$, for $k_{max}
\propto N^{1/\tau}$.

\subsection{High Degree Nodes and the Giant Connected Component}\label{conc}
Let us address the question of which nodes are most likely to be in the giant
connected component when $\beta =\displaystyle\rho^{-1} \frac{\tau -2}{3-\tau}$ is in the order of one. Note that
the percolation probability is $q_c =\langle k \rangle/(\langle k^2
\rangle-\langle k \rangle )\propto k_{max}^{\tau-3}$ for large
$k_{max}$, when $2\leq\tau\leq 3$. As such, the probability that
any node with $k$ links will have any edges left after
percolation (with $\beta$ in the order of one) is around $\sim k
k_{max}^{\tau-3}$. For this probability not to be negligible, we
must have $k\sim k_{max}^{3-\tau}$ or greater. Put in other words, take any
node with degree $k_{max}^{3-\tau-\epsilon}$ for any finite
$\epsilon>0$, then with probability one this node will lose all
its edges after the bond percolation and will not be in the
largest connected component after percolation at a multiple of the
threshold.

\subsection{Percolation Search}
The percolation search algorithm, developed by the authors, is
based on a probabilistic broadcasting scheme, with as small a
probability of broadcast as possible. The success of this
algorithm depends on finding a percolation probability which would
result in \emph{most} of the high-degree nodes to fall into one
giant connected component, while as few \emph{low-connectivity}
nodes would be present in the same component. This will ensure
that most of the search traffic will be carried out by high
capability nodes that have assumed large degrees, while the low
connectivity nodes will only \emph{occasionally} participate in a
search. Moreover, since a query message will be passed only along
a few edges, the protocol will result in low overall traffic. The
results in this paper (see the preceding discussions) show that
random power-law networks are ideally suited for percolation
search, and broadcasting with probability just above the
percolation threshold leads to high query hit rates. For more
details see \cite{perc}.
\bibliography{PrefAttack}

\begin{appendix}

\section{Case of $\tau=2$}\label{ap:tau2}
In the case of $\tau = 2$, (Eqn. \ref{eq:z_alpha}) reads as:
\begin{equation}
\label{eq:z_2} -\alpha z = \sum_{n=1}^\infty \frac{(-z)^{n}}{n!n}
\end{equation}

The above series is related to the $Ei(z)$, or exponential
integral function:
\begin{eqnarray*}
Ei(z) &=& - \int_{-z}^\infty \frac{e^{-t}}{t}dt \ .
\end{eqnarray*}
The series representation of $Ei(z)$ is:
\begin{eqnarray*}
Ei(z) &=& \sum_{n=1}^\infty \frac{z^{n}}{n!n} + \gamma + \ln|z|
\end{eqnarray*}
Where $\gamma$ is the Euler-Mascheroni constant and is
approximately $0.5772$. This gives us:
\begin{equation}
\label{eq:series_2} \sum_{n=1}^\infty \frac{(-z)^{n}}{n!n} =
Ei(-z) - \gamma - \ln|z|
\end{equation}
We can then write (Eqn. \ref{eq:z_2}) as:
\begin{equation}
\label{eq:alpha_z_2} -\alpha z = Ei(-z) - \gamma - \ln z
\end{equation}
when $\tau=2$. For $ z \gg 1$, $Ei(-z)\rightarrow 0$, or
$z/\ln(z)\approx \alpha^{-1}=\rho$ or equivalently, $z\approx \rho
\ln (\rho)$.\vspace{0.1in}
\section{Verifying the percolation threshold}\label{ap:perc}
\begin{figure}
\begin{center}
\includegraphics[width=3in,height=3.3in,angle=270]{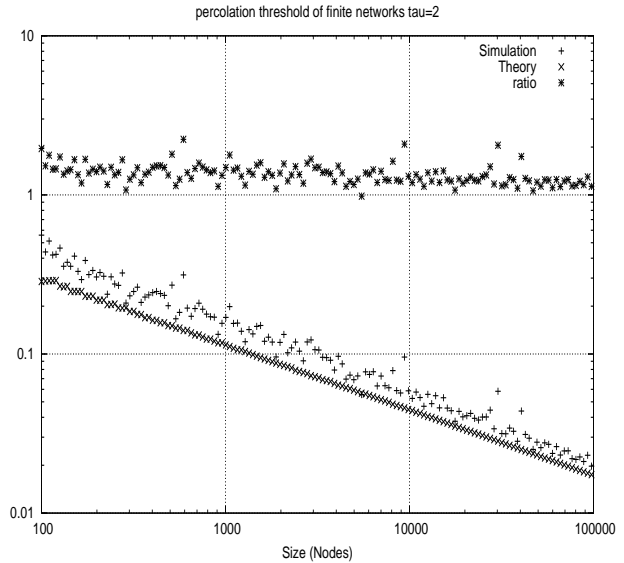}
\caption{ For a finite network we define $S(q)$ as the size of the
largest component as function of percolation probability $q$ and
find the value of $q$ where $dS/dq$ is maximized (marked by the
Simulation points). We compare that value with the percolation
threshold of an infinite network with the same degree
distribution: $q_c = \ave{k}/(\ave{k^2} - \ave{k})$.  We see that
the ratio of these two values is approximately constant and tends
towards unity as the network grows in size. }
\label{fig:threshold}
\end{center}
\end{figure}

The discussions in this paper highlighted the significance of
$q_c(N)= \langle k  \rangle/(\langle k^2  \rangle-\langle k
\rangle)$ as the true finite size percolation threshold. Strictly
speaking, however, $q_c(N)>0$ (for any fixed $N$) is the percolation threshold of an infinite
size random graph whose degree distribution is given by the
truncated PL: $p_k = A_{\tau}
k^{-\tau}$ for $k\in (1, k_{max})$, and $p_k=0$ for $k>k_{max}$ (we let $k_{max}$
scale as $N^{1/\tau}$, and hence is fixed for a given $N$).  We know from percolation theory \cite{MR95,Aiello00} that for a random graph with  a fixed
$k_{max}$, and the degree distribution satisfying certain additional technical conditions, the infinite size percolation threshold can be
shown to be the limit of a uniformly convergent series, and that
the phase transition for any network of finite but large size, $M$, will
also happen close to the infinite size percolation threshold.
The exact dependence of the network size $M_0$ (above which the
infinite size percolation threshold will be a good approximation for finite
size percolation) on the nature of the degree distribution and
$k_{max}$ (especially when it is large) is currently unknown.  In this paper, we are assuming that
for any fixed but large $N$ (which determines $k_{max}$, and hence $q_c(N)$, for
the family of random networks),
the infinite size percolation threshold is also a good approximation
for a finite network of size $N$.

Figure \ref{fig:threshold} directly compares what we might call
the empirical threshold to the infinite network model. If $S(q)$
is the size of the largest component for a fixed network size with
a bond percolation probability $q$, we define the empirical
threshold as the point where $dS(q)/dq$ is maximized. For all
networks, the empirical threshold is very close to the infinite
network model, and as $N$ grows, these two agree more closely.

\end{appendix}
\end{document}